\documentclass[twocolumn,showpacs,preprintnumbers,amsmath,amssymb,A4paper]{revtex4}

\usepackage[dvips]{graphicx}
\usepackage{dcolumn}
\usepackage{bm}

\begin{document}
\newcommand{\kp}{{\bf k$\cdot$p}\ }

\preprint{APS/123-QED}
 \title{Electrons in finite superlattices: the birth of crystal momentum}

\author{P. Pfeffer$^*$ and W. Zawadzki}
 \affiliation{Institute of Physics, Polish Academy of Sciences\\
 Al.Lotnikow 32/46, 02--668 Warsaw, Poland\footnotetext{$^*$ e-mail address: pfeff@ifpan.edu.pl}
  \\}

\date{\today}

\begin{abstract}
Properties of electrons in superlattices (SLs) of a finite length are described using standing waves resulting
from the fixed boundary conditions (FBCs) at both ends. These electron properties are compared with those
predicted by the standard treatments using running waves (Bloch states) resulting from the cyclic boundary
conditions (CBCs). It is shown that, while the total number of eigenenergies in a miniband is the same according
to both treatments, the number of \emph{different} energies is twice higher according to the FBCs. It is also
shown that the wave vector values corresponding to the eigenenergies are spaced twice as densely for the FBCs as
for the CBCs. The reason is that a running wave is characterized by a single value of wave vector $k$, while a
standing wave in a finite SL is characterized by a pair of wavevectors $\pm q$. Using numerical solutions of the
Schroedinger equation for an electron in an increasing number $N$ of periodic quantum wells (beginning with $N$
= 2) we investigate the "birth" of an energy miniband and of a Brillouin zone according to the two approaches.
Using the Fourier transforms of the computed wave functions for a few quantum wells we follow the "birth" of
electron's momentum. It turns out that the latter can be discerned already for a system of \emph{two} wells. We
show that the number of higher values of the wave vector $q$ involved in an eigenenergy state is twice higher
for a standing wave with FBCs than for a corresponding Bloch state. Experiments using photons and phonons are
proposed to observe the described properties of electrons in finite superlattices.
\end{abstract}

 \maketitle

\section{\label{sec:level1}INTRODUCTION\protect\\ \lowercase{}}

Semiconductor superlattices (SLs) have been, since their creation in the early seventies [1], a subject of
intensive studies because of their inherent scientific interest as well as important applications. Together with
other heterostructures, SLs belong to "hand made" quantum systems which can be used to study fundamental
features of quantum mechanics. In particular, SLs are good examples of periodic or quasiperiodic structures
exhibiting quantum effects of periodicity. From their beginning, SLs have been treated theoretically by methods
developed earlier for atomic crystals. Thus, the standard notions of Bloch functions, energy bands, forbidden
gaps, crystal momenta, Brillouin zones, etc., have been used for their description, see [2]. It is true that a
SL and an atomic crystal exhibit many similarities. The main practical difference between them is that a typical
crystal contains millions of unit cells, while a one-dimensional SL may contain only tens or hundreds of quantum
wells (QWs). This means that the usually imposed cyclic boundary conditions (CBCs) apply much better to an
atomic crystal than to a SL. Also, the notions of continuous energy bands and wave vector spaces apply much
better to a crystal than to a SL.

The purpose of our present work is threefold. First, we explore to what extent the notions taken from "infinite"
atomic crystals apply to finite superlattices. In particular, we focus on periodic systems consisting of a few
quantum wells and analyze how the features characteristic of long periodic systems are "formed" as the number of
QWs increases. Second, we work out specific features of electrons in finite SLs and, third, we propose
experiments to observe them. The essential point of our approach is that we realistically consider a finite
superlattice to be a quantum well. Concentrating on important features we consider only the ground energy
miniband, we do not treat higher minibands and states in energy gaps.

The paper is organized as follows. In Section II we review very briefly properties of "infinite" superlattices
(or crystals) with imposed CBCs. Next we consider theoretically various aspects of finite SLs. Finally, we
discuss our results and propose experiments to observe the described new features. The paper is concluded by a
summary.

\section{\label{sec:level2}"INFINITE" SUPERLATTICES WITH CYCLIC BOUNDARY CONDITIONS\protect\\ \lowercase{}}

In this section we consider very briefly "infinite" supperlattices and crystals using CBCs. Since electrons in
"infinite" crystals are described in many textbooks, we only emphasize their main features treating them as a
starting point for our main considerations. An "infinite" sequence of quantum wells is sketched schematically in
Fig. 1, which also indicates parameters of the SL rectangular potential. The potential energy $V(z)$ is a
periodic function of $z$ with the period $d$:
\begin{equation}
V_{SL}(z)=\sum^{+\infty}_{l = -\infty} V(z-ld) \;\;,
\end{equation}
in which

\[V(z-ld) =\left\{\begin{array}{ll} -V_b & \mbox{if $ |z-ld|\leq a/2$} \\
0 & \mbox{if $ |z-ld| > a/2\;\;.$}\end{array}\right.\] The zero of energy is taken at the barrier's height.
Exploiting the periodicity of $V_{SL}(z)$ one considers the translation operator $\tau$ which is such that for
any function $f(z)$ there is ${\tau}f(z) = f(z+d)$. The operator $\tau$ commutes with the Hamiltonian $H$ of the
electron because of the periodicity of $V(z)$. One can thus find the eigenfunctions of $H$ which are
simultaneously the eigenfunctions of $\tau$. These common eigenfunctions are the Bloch states
\begin{equation}
\chi_k(z)=e^{ikz}u_k(z) \;\;,
\end{equation}
in which $u_k(z)$ is a periodic function possessing the periodicity of the potential. It is usually assumed that
the length of the crystal $L = Nd$ is very large and one can impose the cyclic Born-von Karman boundary
conditions on the eigenstates of $H$. Then $\chi_k(z)$ obey
\begin{equation}
\chi_k(z+Nd)=\chi_k(z) \;\;,
\end{equation}
which, upon the periodicity of $u_k(z)$, gives: $e^{ik(z+Nd)} = e^{ikz}$, so that $e^{(ikNd)} = 1$, and the well
known quantization of $k$ is obtained

\begin{equation}
k_j = \frac{2\pi j}{Nd} = \frac{2\pi}{L}j \;\;,
\end{equation}
where $j = 0, 1, 2,...N$. There are $N$ independent values of $j$, the spacing between any two consecutive $k$
values being $2\pi/L$. For $j = N$ we have $k_N = 2\pi/d$. Without loss of generality one may restrict $k$ to
the range [-$\pi/d, +\pi/d$] which is called the first Brillouin zone (BZ). Then $j = 0, \pm 1, \pm 2,...\pm
N/2$. The quantity $\hbar k$ is called crystal momentum, for the $k$ values within the first BZ it plays the
role of momentum. If a crystal lattice is characterized by an inversion symmetry, the electron waves with $k$
and $-k$ are equivalent and the energy has the property ${\cal E}(-k) = {\cal E}(k)$, see e.g. [3]. Thus each
energy in a Brillouin zone has a double degeneracy (without spin). One proves that ${\cal E}(k)$ relation for a
given band is periodic in the $k$ space: the ${\cal E}(k)$ relation in the first BZ [$-\pi/d, +\pi/d$] is
repeated in the second BZ [$+\pi/d, +3\pi/d$], etc., and similarly for the negative $k$ values, see below.

\begin{figure}
\includegraphics[scale=1.1,angle=0, bb =20 25 220 80]{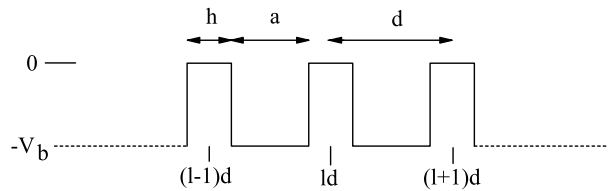}
\caption{\label{fig:epsart}{Rectangular periodic potential of a superlattice used in the calculations: a)
Imposing cyclic boundary conditions, see Eq. (3); b) Imposing fixed boundary conditions in the presence of thick
barriers at both ends, see Eq. (6).}} \label{fig1th}
\end{figure}

Two remarks are in order. First, the precise values of $k_j$ (including $k_0$ = 0) are allowed because the
electron in an "infinite" crystal is completely delocalized, so its momentum may be given exactly. Second, it is
seen from Eq. (2) that the Bloch states represent running waves. Again, this is possible because the crystal is
assumed to be infinite, so the electron can propagate in one direction and does not bounce back. For an
"infinite" rectangular SL the solution of the one-dimensional Schroedinger equation is known exactly for both
well-acting and barrier-acting layers. If the electron effective mass is the same in the wells and in barriers
(which is an idealization) both the wave function and its derivative must be continuous at the two inequivalent
interfaces. Using these boundary conditions one obtains four equations from which a relation between the wave
vector $k$ and the energy $\cal E$ is obtained in the form (cf. [4], [5])

$$
cos(k_j d)=cos(k_w a)cosh(\kappa_b h)+
$$
\begin{equation}
-\frac{1}{2}\left[\frac{-\kappa_b}{k_w}-\frac{k_w}{\kappa_b}\right]sin(k_w a)sinh(\kappa_b h)
 \;\;,
\end{equation}
where $\kappa_b = \sqrt{-2m^*{\cal E}/\hbar^2}$, $k_w = \sqrt{2m^*({\cal E}+V_b)/\hbar^2}$. Relation (5) is
valid for the energies $-V_b \leq {\cal E}(k_j) \leq 0$. In the limit of very thick barriers the RHS of Eq. (5)
goes over to the ${\cal E}(k)$ relation for a single well. Equation (5) gives all energy levels ${\cal E}(k_j)$
corresponding to both even and odd states. The energy minibands of negative energies result from a hybridization
of isolated well states due to coupling across finite barriers. The energies ${\cal E}(k_j)$ calculated from Eq.
(5) for the ground energy miniband are quoted below.

\section{\label{sec:level3}FINITE SUPERLATTICES \protect\\ \lowercase{}}

\begin{figure}
\includegraphics[scale=0.85,angle=0, bb = 75 30 245 220]{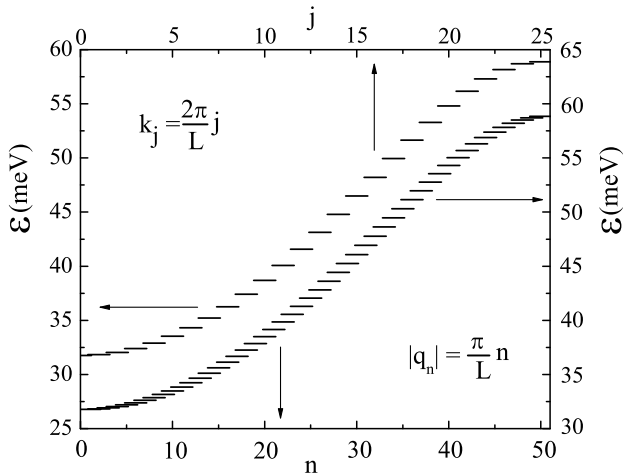}
\caption{\label{fig:epsart}{Theoretical ground energy miniband of a superlattice consisting of $N = 50$ QWs.
Upper trace: result of standard calculation based on CBCs and running waves [see Eq.(5)]. The other half of the
miniband corresponds to negative $j$ values (not shown). Lower trace: result of present calculation based on
FBCs and standing waves. The upper trace is shifted 5 meV upwards for clarity. The values of wave vectors in the
two cases are indicated by the formulas.}} \label{fig2th}
\end{figure}

We now turn to our main considerations regarding the electron behavior in finite superlattices. In order to
concentrate on main features we take a simple rectangular potential, similar to the one shown in Fig. 1, and
assume again that the electron effective mass is the same in the wells and in barriers. The essential new
feature is that the SL is made finite by putting thick barriers on its both sides having the same height as the
barriers between quantum wells (QWs). The total length of our SL is $L = Nd$. We intend to compare the resulting
electron properties with those mentioned in the previous section for "infinite" SL. For a finite superlattice we
are not able to derive analytical results, so our conclusions will be mostly illustrated by numerical
calculations. In all our computations we use the values $a = 8$ nm, $h = 2$ nm, $d = 10$ nm, $V_B = 240$ meV
(see Fig. 1), and the effective mass $m^*$ = 0.067 $m_0$ (the same for the wells and barriers). The considered
structure is characterized by the inversion symmetry. The Schroedinger equation is solved numerically using the
potential shown in Fig. 1, for a given number $N$ of QWs. The boundary conditions at each interface assure the
continuity of the wave function and its derivative.

As to the electron behavior, there exist essential differences between the standard model of an "infinite"
crystal (or superlattice) and our finite superlattice. First, in our case {\emph{the potential is not periodic}}
because at each point $z$ the distances from both ends vary. Second, in our case the electron wave function may
\emph{not} be represented by a \emph{running wave} [see Eq. (2)], but it must be a \emph{standing wave} of some
sort, characteristic of a finite height quantum well. In classical terms, in an infinite periodic structure the
electron with the crystal momentum $\hbar k$ (or $-\hbar k$) runs indefinitely in one (or the other) direction
without coming back. In a finite structure it bounces back and forth which means that in each state both $\hbar
q$ and $-\hbar q$ values are simultaneously involved. Third, since in a finite SL the electron is confined to a
length of about $L$, the value of its \emph{momentum may not be precisely given} because of the uncertainly
principle. Since $\Delta z \approx L$ the uncertainly $\Delta p$ must go as $1/L$. Numerical calculations
confirm this result, see below.

\begin{figure}
\includegraphics[scale=0.90,angle=0, bb = 70 25 240 200]{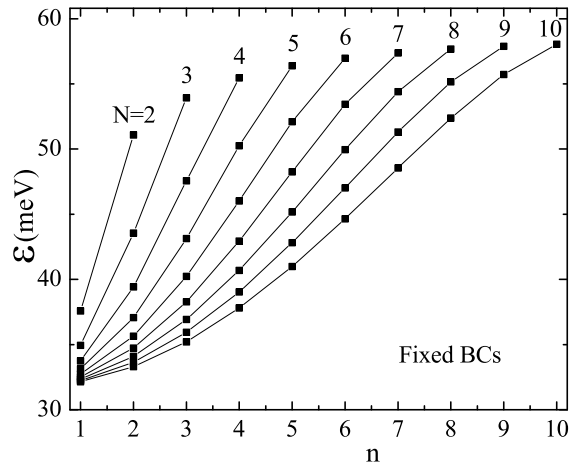}
\caption{\label{fig:epsart}{Low energies of electrons in SLs with increasing number $N$ of QWs versus wave
vector $n$ (in $\pi/L$ units), as calculated using fixed BCs. A formation of s-like shape of a miniband is seen.
The lines join the points to guide the eye.}} \label{fig3th}
\end{figure}

\begin{figure}
\includegraphics[scale=0.90,angle=0, bb = 70 25 240 200]{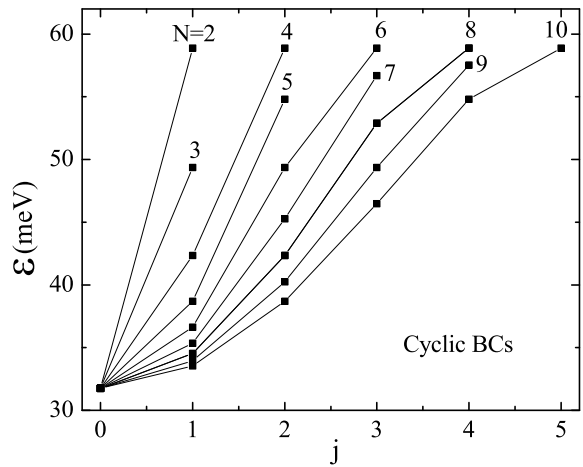}
\caption{\label{fig:epsart}{Low energies of electrons in  SLs with increasing number $N$ of QWs versus wave
vector $j$ (in $2\pi/L$ units), as calculated using cyclic BCs. A formation of s-like shape of a miniband is
seen. The lines join the points to guide the eye.}} \label{fig4th}
\end{figure}

We first consider eigenvalues of the energy which, in contrast to momentum, is a good quantum number for our
system. Figure 2 shows the computed electron energies in the ground miniband for a SL of $N = 50$ QWs. These
eigenenergies are compared with those calculated from Eq. (5) for an SL with the cyclic boundary condition for
$N = 50$. It is seen that one obtains in both cases a similar ground miniband with the same energy width. Also,
the \emph{total} number of allowed energies is the same. On the other hand, the number of \emph{ different}
allowed energies for a finite SL is \emph{twice} that for an SL with CBCs and it is equal to the number of QWs.
This is due to the fact that for an "infinite" SL each energy is twice degenerate with respect to $k$ and $-k$,
while for a finite SL this degeneracy does not occur. This result is in agreement with the well known fact that
for two QWs the energies of the symmetric and odd electron states are not the same. A closer inspection of Fig.
2 indicates that, for a finite SL, one does not deal with split doublets of the degenerate energies. Instead,
the energies are distributed rather uniformly.

Now we estimate quantized values of the quantity $q$ (we hesitate as yet to call it crystal momentum). The wave
functions in a finite SL are standing waves, they should satisfy the following approximate fixed boundary
conditions (FBCs)
\begin{equation}
\Psi(-L/2) = \Psi(L/2) \approx 0 \;\;.
\end{equation}

\begin{figure}
\includegraphics[scale=0.9,angle=0, bb = 55 25 240 200]{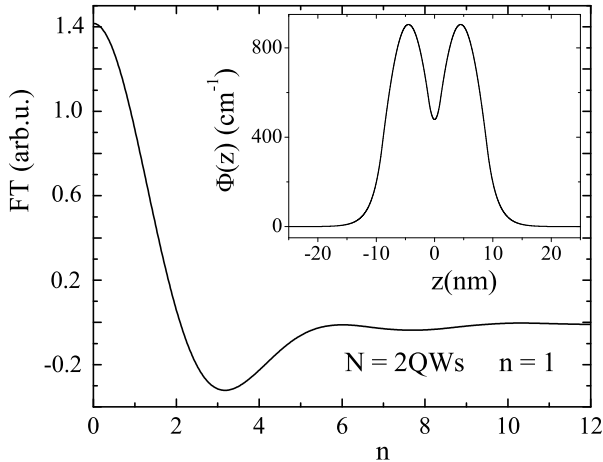}
\caption{\label{fig:epsart}{Calculated Fourier transform of the even wave function (lower energy state, $n$ = 1)
for $N = 2$ QWs versus wave vector $n$ (in $\pi/L$ units). The wave function is shown in the inset.}}
\label{fig5th}
\end{figure}

In principle, our problem is that of an electron in a finite-height rectangular well but for simplicity we will
approximate it by considering an infinitely high rectangular well. Then the eigenfunctions are either $\Psi(z) =
A cos(qz)$ (even states), or $\Psi(z) = B sin(qz)$ (odd states). The FBCs for the cosine function give  $qL =
(1+2r)\pi$, where $r = 0, \pm 1, \pm 2, ...,$ while the FBCs for the sine function give $qL = 2r\pi$, where $r =
\pm 1, \pm 2, ....$ The above conditions may be combined into a simple equation
\begin{equation}
|q_n| = \frac{\pi}{L}n \;\;,\;\;\;\;\;\;n = 1, 2, 3,....
\end{equation}

It is seen that the values of $q_n$, as given by Eq. (7) for a finite SL and resulting from FBCs, are spaced
\emph{twice as densely} as those for an "infinite" structure according to CBCs. Still, the total number of wave
vector states in both cases is the same and is equal to $N$. The difference is that for CBCs the states
correspond to \emph{single} $k$ values: $k = (2\pi/L)(-N/2),...,(2\pi/L)(N/2)$, while for FBCs the states
correspond to \emph{pairs} of values: $q = \pm (\pi/L)1,...,\pm(\pi/L)N$.

The energies ${\cal E}$ for "infinite" and finite SLs are plotted as functions of $j$ and $n$ in Fig. 2.
According to the above considerations, the upper trace in Fig. 2 describes half of the BZ according to CBCs (the
other half is given by $-25 \le j \le 0$), while the lower trace corresponds to the complete BZ according to
FBCs. We emphasize again that the relation (7) is approximate, as it has been obtained from the consideration of
an infinitely high rectangular well. We come back to this point below. It should be noted that the CBCs allow
for $k_0 = 0$, while for the FBCs the value $q = 0$ is not allowed, see the discussion below.

 A one-dimensional density of states in the energy space is $\rho({\cal E}) \sim dk/d{\cal E}$.
It is clear that, if $\rho({\cal E})$ is considered in an energy range including many levels, it will be similar
(if not identical) in both cases because the ${\cal E}(k)$ and ${\cal E}(q)$ relations are similar. However, if
one is interested in $\rho({\cal E})$ on the scale of one or two levels, the two cases will give considerably
different results.

\begin{figure}
\includegraphics[scale=0.9,angle=0, bb = 55 25 240 200]{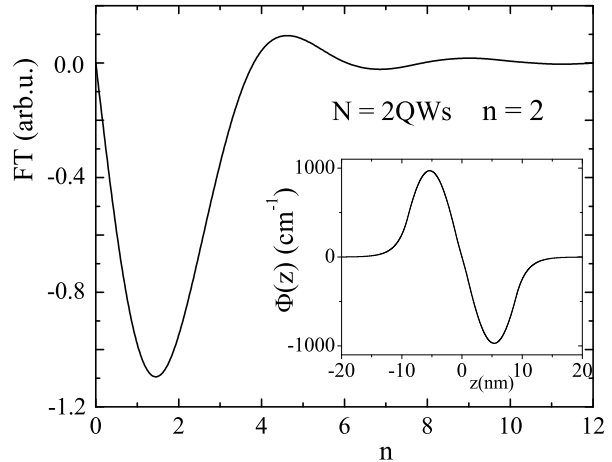}
\caption{\label{fig:epsart}{Calculated Fourier transform of the odd wave function (upper energy state, $n$ = 2)
for $N = 2$ QWs versus wave vector $n$ (in $\pi/L$ units). Strong peak below $n =2$ is seen. The wave function
is shown in the inset.}} \label{fig6th}
\end{figure}

It is of interest to study how the energy band shown in Fig. 2 is formed. We consider it first for the FBCs.
Clearly, for one QW one can not have a band since there is no periodicity. For two QWs, the two lowest energies
(resulting from the above mentioned splitting related to the even and odd states) already form a "germ" of the
ground energy band. When more QWs are added, the energies gradually develop into a real band. This process is
illustrated in Fig. 3, in which we show the calculated energies for nine SLs consisting of $N = 2$ to $N = 10$
QWs. It is seen that, beginning with $N = 4$, the energy values ${\cal E}(q_n)$ form the characteristic s-like
shape of an energy band. It is also seen that the width of the ground energy band $\Delta{\cal E}$ increases
with increasing $N$. However, the width $\Delta{\cal E}$ quickly saturates, so that for $N = 10$ it almost
reaches its final value. Thus for $N = 10$ the energy width shown in Fig. 3 is 25.9 meV, while the width shown
in Fig. 2 for $N = 50$ is 27.06 meV.

Also, Fig. 3 partially elucidates a formation of the Brillouin zone. One can define a Brillouin zone in various
ways, the simplest would be to say that a BZ is well formed if the ${\cal E}(q_n)$ relation is horizontal at the
point $|q_N| = \pi NL = \pi/d$, so that the electron velocity vanishes at this point: $v = (d{\cal E}/d\hbar
q)_{q_N} \approx 0$.  It is seen from Fig. 3 that, according to this criterion, even for $N = 10$ the dependence
${\cal E}(q_n)$ is not horizontal at $q_{10}$. In other words, a SL of 10 QWs is not "periodic enough" to form a
good Brillouin zone.

Now we consider the same problems according to the CBCs, i.e. using Eq. (5) for the eigenenergies. This approach
is clearly not good for few QWs but we try it anyway in order to expose its limitations. The results are shown
in Fig. 4 for an increasing number $N$ of QWs. It follows from Fig. 4 that for the extreme case of $N = 2$, the
model gives \emph{three} energies: ${\cal E}_{\pm1}$ and ${\cal E}_0$. The last one could be considered as an
artefact but, interestingly, the energy difference ${\cal E}_1 - {\cal E}_0$ gives the exact width of the
miniband reached at high values of $N$. It is seen that, for even $N$ values, the number of eigenenergies is $N
+ 1$ and the miniband width is always the same. For $N = 3$ there are also three eigenenergies: ${\cal
E}_{\pm1}$ and ${\cal E}_0$. For odd $N$ values the number of eigenenergies is always $N$ and the width of a
miniband increases with $N$. In that sense the model of CBCs is more "natural" for odd $N$ values. As expected,
the exact calculation for a few QWs with the use of FBCs is distinctly better than that using CBCs, while for
high $N$ values the two models give similar minibands, as illustrated in Fig. 2.

Finally, we consider the problem of whether the electron in a finite SL can be characterized by a crystal
momentum. To put the question differently: how many periodic quantum wells does one need to be able to talk of
the crystal momentum ? Trying to answer this question with the use of FBCs we first compute the wave functions
corresponding to specific eigenenergies and then calculate their Fourier spectra to see what values of the wave
vector are involved in them. It is clear without any calculations that, since we are dealing with standing waves
and the system is characterized by the inversion symmetry, the Fourier transforms (FTs) must be symmetric in $q$
and $-q$. To be more specific, we begin with a SL of $N = 2$ and calculate the wave functions corresponding to
the eigenenergies shown in Fig. 3. The lowest state is even ($n = 1$) and the higher state is odd ($n = 2$).
Their FTs are shown in Figs. 5 and 6. For $n =1$ the maximum of FT is at $q_1 = 0$. This result is not
reasonable because one expects the maximum to be near $|q_1| = 1$ (in $\pi/L$ units). The reason is that for $n
= 1$ the corresponding wavelength is $\lambda_1 = 2\pi /q_1 = 2L$. Since the length of our SL is $L$, it follows
that for $n = 1$ only half of the full wavelength fits into the SL. It is known that one needs at least one full
wavelength to fit into the considered length to make the Fourier analysis meaningful. This is the case for $n =
2$, because $\lambda_2 = 2\pi /q_2 = L$, so that one full wavelength fits into the SL. It is seen from Fig. 5
that the FT for the upper energy state has a pronounced extremum below $n = 2$.

The above results merit some comments. First, the reasoning for $n = 1$ is equally valid for any $N$, so that
also for longer SLs the FTs for $n = 1$ are not meaningful. Our calculations confirm this conclusion. Second, it
is seen that, as expected, the main peak has a finite width $\Delta n$. Still, anticipating a little we can say
that already for $N = 2$ the electron motion in the state $n = 2$ is dominated by a specific value of crystal
momentum (positive and negative).

\begin{figure}
\includegraphics[scale=0.93,angle=0, bb = 55 20 240 240]{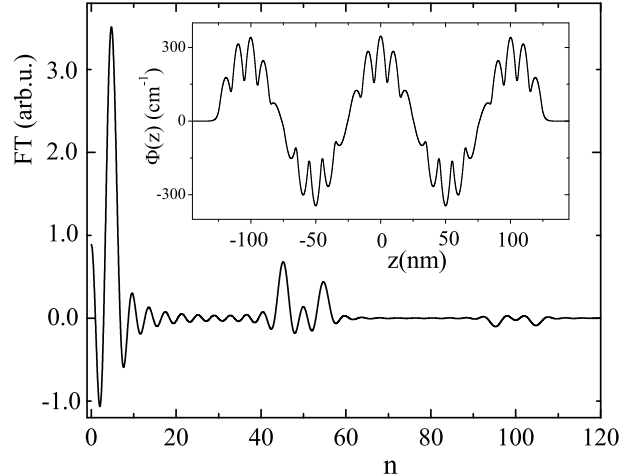}
\caption{\label{fig:epsart}{Calculated Fourier transform of the wave function (shown in the inset) for $N = 25$
QWs and $n = 5$ versus wave vector $n$ (in $\pi/L$ units). The basic peak just below $n = 5$ and high-$n$ peaks
are seen. The wave function is a product of the envelope (standing wave) and the periodic component.}}
\label{fig7th}
\end{figure}

Clearly, SLs with higher number of QWs possess a better pronounced periodicity. In the insets of Figs. 7, 8 and
9 we show, as a matter of example, calculated wave functions corresponding to the eigenenergies ${\cal E}_n$ for
SLs of $N = 25, 50$ and 100 QWs, respectively. They resemble the Bloch states by having quickly oscillating
periodic components and slowly varying envelopes. Still, in true Bloch states the envelope functions are running
waves $exp(\pm ikz)$, while in our case the envelope functions are standing waves of $sin(qz)$ and $cos(qz)$
type.

The Fourier transforms of the above wave functions are shown in Figs. 7, 8 and 9, they indicate what crystal
momenta are involved in the corresponding electron motion. We first consider Fig. 7. It is seen that we deal
with a pronounced peak corresponding approximately to the "main" value of crystal momentum $n = 5$. The peak has
the width of $\Delta n \approx 4$. Since the SL has the length $L \approx \Delta z$, it follows that $\Delta z
\cdot \hbar\Delta q \approx L \hbar 4\pi/L = 4\pi \hbar$. Thus the width $\Delta q$ is directly related to the
uncertainty principle. Our calculations for SLs of different numbers $N$ show that the main peak has always
about the same width $\Delta n \approx 4$. The corresponding width $\Delta q \approx 4\pi/L$, so that the
uncertainty of momentum decreases when the length of SL increases. This means that for really long SLs the main
FT peak approaches the Dirac $\delta$-function. We emphasize again that the complete Fourier transforms contain
not only the functions shown in Figs. 6, 7, 8 and 9 but also their mirror images for negative $n$ (and $q$).

\begin{figure}
\includegraphics[scale=0.93,angle=0, bb = 55 20 240 240]{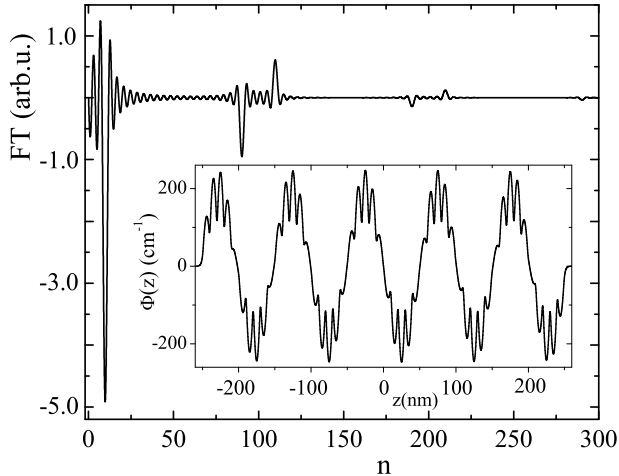}
\caption{\label{fig:epsart}{The same as in Fig. 7, but for $N = 50$ QWs and $n = 10$. The basic peak just below
$n = 10$ and high-n peaks are seen.}} \label{fig8th}
\end{figure}

The remarkable feature seen in the FTs shown in Figs. 7, 8 and 9 are additional sharp peaks at higher $n$
values. In Fig. 7, in addition to the main peak near $n = 5$, there are peaks at $n = 45, 55, 95, 105,...$. In
Fig. 8, the main peak is near $n = 10$, the additional peaks occur at $n = 90, 110, 190, 210,...$. The origin of
the additional peaks is understood when one recalls the fact well known from the theory of infinite periodic
systems, that the ${\cal E}(k)$ relation for each energy band is a periodic function of $k$ (see e.g. Ref. 3).
We plot schematically a similar function for a finite SL, see Fig. 10. It is seen that a given value of the
energy corresponds to many possible values of $n$ (or $q$) and the additional peaks occur exactly at these $n$
values. We demonstrate this rule considering the results shown in Fig. 7. The main peak is at $n_1 = 5$ and the
edge of the BZ is at $N = 25$. According to Fig. 10 the horizontal line at the energy ${\cal E}(n_1)$ crosses
the ${\cal E}(n)$ dependence first at $n_2 = N-n_1 +N = 45$, then at $n_3 = 2N+n_1 = 55$ etc, in agreement with
the above values. This result can be simply interpreted recalling that to a given eigenenergy ${\cal E}_n$ there
always corresponds predominantly a pair of $-q_n$ and $+q_n$ wave vectors in the first BZ. Then the additional
peaks are simply repetitions of the above fundamental pair in the second BZ, third BZ, etc. Not knowing anything
about the theory of periodic structures but only regarding the function shown in Fig. 7, one would expect its FT
to reflect a long wavelength (low $n$) due to the envelope and a short wavelength (high $n$) due to the periodic
component. The latter contains 25 oscillations in Fig. 7, 50 oscillations in Fig. 8, and 100 oscillations in
Fig. 9. What one finds in Figs. 7, 8 and 9 are the expected main peaks at $n_1$, while the high-$n$ peaks are
combinations of $N$ with $n_1$. Comparing heights of the main and high-$n$ peaks in Figs. 6, 7, 8, 9 one sees
that they increase with $N$ and the peaks resemble more and more the $\delta$-functions. The reason is that in
longer SLs the electrons are less localized and their momenta can be specified with greater precision. Also,
with increasing $N$ the main peaks coincide better with the corresponding values of $n$. It is because of this
coincidence that we could associate the eigenenergies with specific values of $n$ in Fig. 2. All in all, our
considerations of the FTs can be summarized as follows: 1) the main $n$ peaks are broadened because we deal with
finite-length SLs, 2) high-$n$ peaks reflect the periodicity of the structure. To this we want to add an
important comment that finite SLs are \emph{not} characterized by an equivalence of states with the wavevectors
$q$ and $q + G_i$, where $G_i$ is the reciprocal lattice vector. This follows from Figs. 7, 8 and 9, where the
main $q$ peaks are much higher then others.

\begin{figure}
\includegraphics[scale=0.88,angle=0, bb = 50 30 240 220]{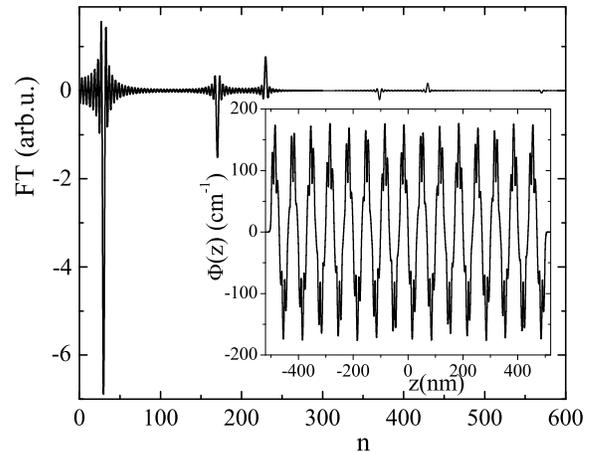}
\caption{\label{fig:epsart}{The same as in Figs. 7 and 8, but for $N = 100$ QWs and $n = 30$. The basic peak at
$n = 30$ and high-n peaks are seen.}} \label{fig9th}
\end{figure}

We can interpret the high-$n$ peaks in the FTs of electron states by recalling a well known property of the
Bloch functions for "infinite" periodic structures [6]. If a Bloch state $\Psi$ contains a certain wave vector
$k_0$, all the wave vectors in the Fourier expansion of this state are given by $k_0 + G_i$, where $G_i$ are the
reciprocal lattice vectors. The FTs shown in Figs. 7, 8 and 9 exhibit exactly this property. The difference
compared to the standard theory is, that a Bloch state is characterized by \emph{one} value $k_0$, whereas our
states are characterized by a \emph{pair} of values $\pm q_0$. In consequence, each value gives rise to a
series: $+q_0 + G_i$ and $-q_0 + G_i$, and we deal with twice as many high-$n$ peaks compared to the standard
Bloch states. One can also say that the high-$n$ peaks, corresponding to the descending parts of the ${\cal
E}(n)$ curve (peak $n_2$ in Fig. 10), provide evidence for the mirror-image parts of FTs (with negative $n$)
which we do not show in our figures.

We conclude this section by two remarks of a more general nature. First, as follows from a comparison of Eq. (7)
with Eq. (4) and of Fig. 3 with Fig. 4, the model of CBCs allows for the value of wave vector $k = 0$, while our
realistic calculation for finite SLs with the use of FBCs does not allow this value, the smallest $q$ being $\pm
1$. The last result is a direct consequence of the uncertainty principle. However, it means that, while
according to CBCs the electron can be at rest, in a real finite periodic structure the electron may not be at
rest and we deal with a kind of "zero point oscillations" for the crystal momentum. Second, it follows from the
above considerations that in a finite unperturbed periodic structure the electron always bounces back and forth,
while according to the cyclic BCs it propagates in one direction. The question arises how to reconcile the two
treatments when the structure is long. This can be done introducing scattering events. If the electron scatters
from time to time, it will not bounce back and forth. In other words, if the mean free path $l$ is shorter than
the SL length $L$, a standing wave is equivalent to a running wave. This justifies the common use of Bloch
states in the scattering theory.

\begin{figure}
\includegraphics[scale=0.98,angle=0, bb = 50 30 220 135]{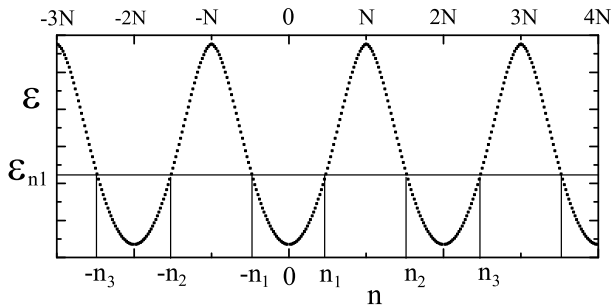}
\caption{\label{fig:epsart}{Periodic ${\cal E}(n)$ function of the ground energy miniband for a SL of $N = 50$
QWs versus $n$ (schematically). Constant-energy line ${\cal E}_{n1}$ intersects the ${\cal E}(n)$ curve in
consecutive Brillouin zones, which determines the high-n components involved in the wave function.}}
\label{fig10th}
\end{figure}

\section{\label{sec:level4}EXPERIMENTAL POSSIBILITIES\protect\\ \lowercase{}}

It appears that experimental observations of characteristic electron properties in finite SLs should not be too
difficult. As we showed above, the number of \emph{different} eigenenergies in a miniband doubles compared to
that in an "infinite" SL described by the CBCs, see Fig. 2. This corresponds to distinctly smaller energy
differences between consecutive eigenstates in a miniband. These energy differences can be directly observed by
infrared optical absorption. A photon carries almost no momentum, so that, if consecutive states were
characterized by sharp different values of the wave vector $q_n$, optical transitions between them would not be
possible. However, as we argued above, the energy eigenstates are characterized by somewhat spread-out $q$
values, so the optical transitions with $q$ conservation are possible. It is clear that for longer SLs, which
are characterized by a "better" periodicity, such optical transitions have a lower probability.

\begin{figure}
\includegraphics[scale=0.85,angle=0, bb = 70 25 240 200]{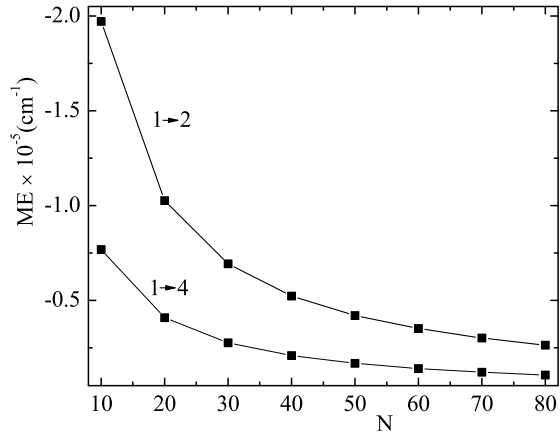}
\caption{\label{fig:epsart}{Matrix elements for optical transitions between 1$\rightarrow$2 and 1$\rightarrow$4
states, calculated for SLs with increasing number $N$ of QWs. The lines join the points to guide the eye.}}
\label{fig11th}
\end{figure}

We calculated the matrix elements for optical transitions in the electric dipole approximation. The Hamiltonian
for the electron-photon interaction is: $H' = (e/m^*c)\textbf{A'}\cdot \textbf{p}$, where $\textbf{A'}$ is the
vector potential of radiation. Thus the matrix elements for optical transitions are $<\Psi_f
|\textbf{p}|\Psi_i>$, where $i$ and $f$ subscripts stand for the initial and final electron states,
respectively. We computed numerically the matrix elements ${<\Psi_2}|{(d\Psi_1/dz)>}$ and
${<\Psi_4}|{(d\Psi_1/dz)>}$ using the electron wave functions of the type shown in Figs. 7, 8 and 9. The results
are shown in Fig. 11 for SLs consisting of $N = 10$ to $N = 80$ QWs. It is seen that, in agreement with the
above considerations, the MEs decrease when $N$ increases. Also, the MEs for 1$\rightarrow$4 transitions are
considerably smaller then those for 1$\rightarrow$2 transitions because the overlap of $q$ values is much
smaller for the former. The MEs for odd$\rightarrow$odd and even$\rightarrow$even transitions vanish since the
corresponding wave functions have the same parity.

In order to get an idea about the absolute values of the MEs we compare them with that for the cyclotron
resonance. The ME for a transition between the zeroth and first Landau levels is $\hbar/L_m$, where $L_m =
(\hbar/eB)^{1/2}$ is the magnetic radius. Since in MEs given in Fig. 11 we omitted $\hbar$, we have to compare
their values with $1/L_m$. For $B = 4$ Tesla there is $1/L_m$ = 7.8$\times 10^5$ cm$^{-1}$. Thus the values of
MEs shown in Fig. 11, although somewhat smaller than that for the cyclotron resonance, seem sufficient to make
the corresponding optical transitions observable.

Next we turn to the peaks of higher wave vector $q$ present in the electron wave functions, see Figs. 7, 8 and
9. These peaks should be observable by a resonant absorption of acoustic phonons. If the phonon dispersion is
${\cal E}_{ph}(Q)$, an absorption process will take place when the momentum conservation $q_f - q_i = Q_{ph}$
and the energy conservation ${\cal E}_f - {\cal E}_i = {\cal E}_{ph}$ are satisfied. This can occur for the same
signs of $q_i$ and $q_f$ (which would require lower values of $\Delta q$), as well as for the opposite signs of
$q_i$ and $q_f$ (larger $\Delta q$). The standard electron-phonon interaction via the deformation potential
mechanism occurs for longitudinal phonons, so the latter should propagate along the growth direction of SL. The
broadening of $q$-peaks, seen in Figs. 6, 7, 8, 9, should somewhat relax the momentum conservation which would
facilitate the resonant phonon absorption. We emphasize again that the number of $q$-peaks in the electron wave
functions available for the resonant transitions with phonons is twice as high for finite SLs as for "infinite"
periodic structures. Transitions between electron states in different Brillouin zones, known as the Umklapp
processes, are known in the physics of 3D crystals. They contribute to electron scattering in transport
phenomena and have usually a nonresonant character.

Coming back to finite SLs, it is difficult to imagine transitions between different $q$-peaks for the same
electron energy since they would require excitations having relatively large momentum and vanishing energy. Such
transitions could only participate in second-order excitations (for example in the Raman scattering), where the
transitions to intermediate states require the momentum conservation but not the energy conservation. Also, one
can imagine a resonant optic phonon emission accompanied by electron transitions between the states of different
energy and momenta. Such phonon emission would cause an energy splitting of the upper electron state, similarly
to the effects observed for the Landau levels of an electron in a magnetic field.

\section{\label{sec:level5}SUMMARY\protect\\ \lowercase{}}

We describe electrons in finite superlattices treating the latter realistically as quantum wells and applying to
them the fixed boundary conditions. We find that the electron wave functions are products of standing waves (of
sine and cosine type) and periodic components. In classical terms, this corresponds to electrons bouncing back
and forth. Our description is compared with the standard approach to electrons in periodic structures, based on
the cyclic (Born-von Karman) boundary conditions, which are satisfied by running waves (Bloch states). We find
that, while the total number of eigenenergies in a miniband is the same according to both treatments, the number
of different eigenenergies is twice larger according to FBCs. We also find that the wave vectors corresponding
to the eigenenergies are spaced twice as densely according to FBCs as according to CBCs. An important difference
between the two approaches is that a running wave is characterized by one value of the wave vector $k$, while a
standing wave is characterized by a pair of wave vectors $\pm q$. Considering SLs with increasing number of QWs
we follow formation of basic entities of periodic structures: energy bands, crystal momenta, and Brillouin
zones. We find that one can characterize electron states by the crystal momentum beginning with $N = 2$ QWs, the
s-like shape of a miniband is formed beginning with $N = 4$ QWs, but for a formation of a Brillouin zone edge
more than $N = 10$ QWs are needed. We calculate numerically the wave functions for finite SLs and then their
Fourier transforms for increasing number of QWs in order to determine the predominant crystal momenta $q$
involved in them. We find that the main $q$ peaks are broadened, which is a direct consequence of the
uncertainty principle. In addition to the main $q_n$ peak, the FTs exhibit smaller peaks at $q_n + G_i$ and
$-q_n + G_i$, where $G_i$ are reciprocal lattice vectors. This is similar to the common property of Bloch
states, but in our case the number of high $q$ peaks is twice higher. Finally, we consider experimental
possibilities to observe the described properties of finite SLs with the use of photons and phonons.

\begin{acknowledgments}
We acknowledge the financial
support of Polish Ministry of Science and Higher Education through
 Laboratory of Physical Foundations of Information Processing.
\end{acknowledgments}

\end{document}